# Modeling Thalamocortical Cell: Impact of Ca$^{2+}$ Channel Distribution and Cell Geometry on Firing Pattern

Reza Zomorrodi[1,2], Helmut Kröger[1], Igor Timofeev[2,3*]

[1] Department of Physics, Laval University, Canada
[2] The Centre de recherche Université Laval Robert-Giffard (CRULRG), Laval University, Canada
[3] Department of Anatomy and Physiology, Laval University, Canada

**Running title:**
Model of thalamocortical neuron

**Correspondence:**
*Igor Timofeev, The Centre de Recherche Université Laval Robert-Giffard, Local F-6500, 2601 de la Canardière, Québec, (QC) G1J 2G3, Canada*
**Phone:** 1-418-663-5747 ext. 6396
**Fax:** 1-418-663-8756

Igor.Timofeev@phs.ulaval.ca

**Abstract**

The influence of calcium channel distribution and geometry of the thalamocortical cell upon its tonic firing and the low threshold spike (LTS) generation was studied in a 3-compartment model, which represents soma, proximal and distal dendrites as well as in multicompartmental model using the morphology of a real reconstructed neuron. Using uniform distribution of $Ca^{2+}$ channels, we determined the minimal number of low threshold voltage-activated calcium channels and their permeability required for the onset of LTS in response to a hyperpolarizing current pulse. In the 3-compartment model, we found that the channel distribution influences the firing pattern only in the range of 3% below the threshold value of *total* T-channel density. In the multi-compartmental model, the LTS could be generated by only 64% of unequally distributed T-channels compared to the minimal number of equally distributed T-channels. For a given channel density and injected current, the tonic firing frequency was found to be inversely proportional to the size of the cell. However, when the $Ca^{2+}$ channel density was elevated in soma or proximal dendrites, then the amplitude of LTS response and burst spike frequencies were determined by the ratio of *total* to *threshold* number of T-channels in the cell for a specific geometry.

**Keywords:** Thalamocortical cell, low-threshold spike, T-current, channel distribution

# 1. Introduction

Thalamocortical (TC) neurons at depolarized membrane potential fire in tonic mode, while when released from a hyperpolarizing state and crossing the potential between -65 and -70 mV they generate low-threshold spikes (LTS), accompanied by sodium spikes (Jahnsen and Llinás, 1984a). During slow-wave sleep, TC neurons are hyperpolarized and fire preferentially LTS accompanied by high-frequency spike-bursts, while during paradoxical sleep and likely other activated brain states they fire mainly in the tonic mode (Hirsch et al., 1983, Steriade et al., 1993). Spike-burst elicited during slow-wave sleep precedes and follows a period of at least 100 ms of silence, lasts 5-20 ms and usually consists of 3-5 action potentials (Domich et al., 1986). Extracellular unit recordings demonstrated also the presence of high frequency spike-trains during waking states (Sherman and Guillery, 2002, Bezdudnaya et al., 2006). Both high frequency spike-trains at depolarized potentials and low-threshold spike-bursts elicited by synaptic volleys can have similar shape despite different underlying mechanisms (Rosanova and Timofeev, 2005). The low-threshold calcium current ($I_T$) underlies burst generation in TC relay cells and plays a central role in the generation of synchronized sleep oscillations (Steriade and Deschénes, 1984, Steriade et al., 1993, Steriade et al., 1997, Destexhe and Sejnowski, 2001, Timofeev et al., 2001).

Ascending and descending inputs to thalamic relay cells arrive at different compartments of the dendritic tree (Jones, 1985) and their integration depends on a large set of intrinsic currents. In order to reproduce correctly synaptic integration, numerical simulations need to incorporate the electrically active properties of the dendritic tree. A multi-compartment model is required to consider the effects of dendritic currents. A first step in this regard was carried out by Destexhe and co-workers (Destexhe et al., 1998) who studied relay cell models with a dendritic tree, which incorporated dendritic T-current densities. In that work, a comparison of T-current density recorded from acutely dissociated and relatively intact cells suggested a higher T-current density in the dendrites compared to the soma. In addition, the high distal T-current density predicted in that model led to a number of speculations (Zhan et al., 2000) on relay cell responses to synaptic inputs arriving to distal dendrites. Fluorescent imaging study (Zhou et al., 1997) and patch clamp recordings from soma and proximal dendrites (Williams and Stuart, 2000) suggested a higher density of T-channels on proximal dendrites. However, a high density of T-channels in distal dendrites was not required to reproduce low-threshold spikes in a model (Rhodes and Llinas, 2005). Previously, both, experimental and modeling studies were not efficient in detecting the presence and the amplitude (value) of T-current in distal dendrites. Thus, the distribution of T-channels in the membrane of TC neurons remains to be elucidated.

Since the active property of the dendrites affects synaptic integration and response of the neuron (De Schutter and Bower, 1994b, a, Mainen and Sejnowski, 1996, Koch, 1999, Goldberg et al., 2007), in the present study, the pattern of T-channel distribution was investigated in computational experiments. Given experimental facts that T-channels are distributed unequally over the TC neuron membrane, we examined the different forms of non-uniform distribution in the multicompartment models in order to find the possible effects of channel distribution on the cell responses. For simplicity, we started with 3-compartment model and then we extent the model to a 1267 compartments with realistic morphology of TC cell. We hypothesize that low-threshold response of TC neuron can be correctly reproduced in a multicompartment model with an appropriate form of T-channel density. In addition, we hypothesize the specific form of channel distribution should have physiologically benefits for cell. Here we show that the number of T-channels in the cell has a prime influence on LTS bursting and, that the shape of the distribution becomes important only in a specific range of the total number of T-channels. Since different types of TC cells have different geometrical parameters, we also investigated the influence of the cell geometry on the firing pattern.

# 2. Methods

## 2.1. Model

The electrophysiological responses of thalamocortical cell could reproduce by a set of coupled differential equations with an appropriate set of parameters (Rose and Hindmarsh, 1985, McMullen and Ly, 1988) based on experiments by Jahnsen and Llinás (Jahnsen and Llinás, 1984a, b). In our simulation, we used the Hodgkin-Huxley type model for the TC cell. In order to reproduce the low-threshold response of the TC neuron, the kinetic of the T-current was taken from voltage clamp experiment (McCormick and Huguenard, 1992).

The main feature of a TC electrophysiological responses cell can be reproduced in a single compartment model involving $Na^+$, $K^+$ and T-currents described by Hodgkin-Huxley equations. Figure 1 shows the tonic response and the burst spiking of the modeled TC neuron. Above we proposed the hypothesis that all features of rebound spike-burst behavior can be reproduced in a multi-compartment model, if one takes into account the channel distribution in dendrites. The 3- compartment model simulation of TC neuron responses shown below has been obtained using the NEURON simulating environment (Hines and Carnevale, 1997).

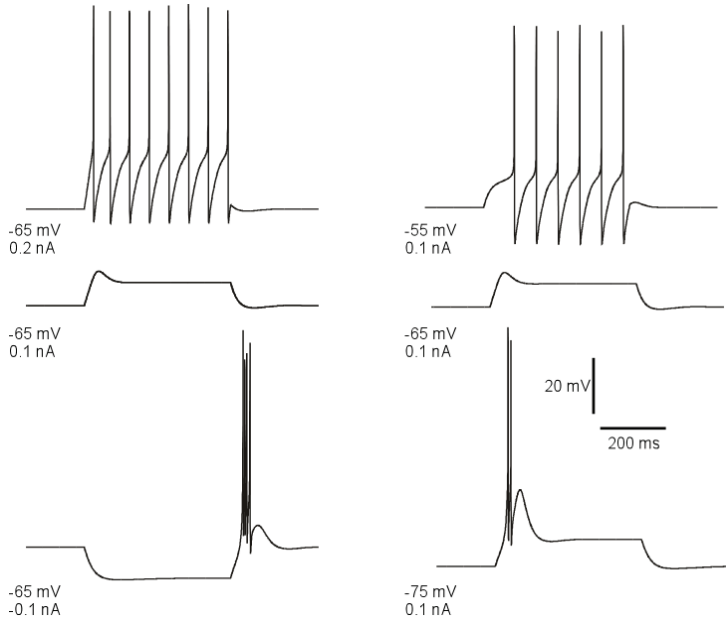

**Figure 1. Simulated rebound burst in TC cell in a single compartment model**.
The model involves the leak current, fast $Na^+/K^+$ currents and the low-threshold $Ca^{2+}$ current. Left panel: response of cell to different injection currents, at the same level of membrane potential: tonic firing for 0.2 nA (top), passive response for 0.1 nA (middle) and burst firing for -0.1 nA (bottom). Right panel: response of the cell to a depolarizing current (0.1 nA) at different levels of membrane potential: tonic firing for $V_m$=-55 mV, passive response for $V_m$=-65 mV and burst firing for $V_m$=-75 mV.

## 2.2. 3-compartment model

The ultimate goal of our study is to understand how T-channel distribution influences electrophysiological responses of real TC neurons. The initial steps of the study were however done in much simpler 3-compartmental model that enables to test multiple parameters with high computational efficiency. The three compartments were the somatic segment and the dendritic arbor composed of proximal and distal segments. According to previous study (Destexhe et al., 1998), the 3-compartmental model provides good precision in reproduction of major electrophysiological features of TC cells. In our simulations, the basic ("control") parameters of the three sections have the following length (*l*) and diameters (*d*): $l_s$=38.41 μm, $d_s$=26 μm for soma, $l_p$=12.49 μm, $d_p$=10.28 μm for proximal section and $l_d$=84.67 μm, $d_d$=8.5 μm for distal section (Fig. 2). It has been assumed that passive parameters (e.g. leak conductance, cytoplasimc resistance, and capacitance) of the cell are constant in each compartment. The passive response of the model with a dendritic correction factor, $C_d$= 7.95 (see Eq. 2.2.1) is fitted to the passive response of the cell during voltage-clamp recording. The Hodgkin-Huxley equations for the three-compartment model are given by

$$(2.2.1)$$

$$C_m \frac{dV_1}{dt} = -g_L(V_1 - E_L) - \bar{g}_{Na} m^3 h(V_1 - E_{Na}) - \bar{g}_k n^4 (V_1 - E_k)$$
$$- \bar{P}_{Ca} r^2 s G(V_1, Ca_o, Ca_i) - g_2^1 \frac{(V_1 - V_2)}{A_1} + I_{inj},$$

$$C_d C_m \frac{dV_2}{dt} = -C_d g_L(V_2 - E_L) - C_d \bar{P}_{Ca} r^2 s G(V_2, Ca_o, Ca_i) - g_2^1 \frac{(V_2 - V_1)}{A_1} - g_3^2 \frac{(V_3 - V_2)}{A_2},$$

$$C_d C_m \frac{dV_3}{dt} = -C_d g_L(V_3 - E_L) - C_d \bar{P}_{Ca} r^2 s G(V_3, Ca_o, Ca_i) - g_3^2 \frac{(V_3 - V_2)}{A_3}.$$

Here $V_1$, $V_2$, and $V_3$ represent the membrane potential of the soma, proximal and distal compartments, respectively. $A_1$, $A_2$, and $A_3$ denote the area of the compartments. In this model, the membrane capacitance, the leak conductance and the leak reversal potential were set to $C_m$ =0.878 µF/cm², $g_L$=0.0379mS/cm², $E_L$=-69.85 mV, respectively. The axial conductance ($g_j^i$) between compartments depends on the geometry of the compartment and the cytoplasimc resistance ($R_a$) in the following way:

$$\frac{1}{g_j^i} = r_j^i = \frac{2 R_a}{\pi} \left( \frac{l_i}{d_i^2} + \frac{l_j}{d_j^2} \right) \quad . \quad (2.2.2)$$

Thus, the axial conductance in our model takes the following values $g^1_2$=5.187 µS and $g^2_3$=0.703 µS. Voltage-dependent conductances were modeled using a Hodgkin–Huxley type of kinetic model. The kinetics of the Na$^+$ and K$^+$ currents responsible for fast action potentials, were taken from a model of hippocampal pyramidal cells (Traub and Miles, 1991), assuming a resting potential of $V_T$= -52 mV, maximal conductance $\bar{g}_{Na}$ =0.1 S/cm² and $\bar{g}_k$ =0.1 S/cm², and reversal potentials of $E_{Na}$ = 50 mV and $E_K$=-100 mV. This model has been shown to be adequate to model the repetitive firing within bursts of action potentials (Destexhe et al., 1996). Because of the nonlinear and far-from-equilibrium behavior of calcium ions to model Ca$^{2+}$ current we used Goldman-Hodgkin-Katz (GHK) equation (Hille, 2001):

$$I_T = P_{Ca} m^2 h G(V, [Ca]_o, [Ca]_i),$$

$$\frac{dm}{dt} = -\frac{1}{\tau_m(V)} (m - m_\infty(V)). \quad (2.2.3)$$
$$\frac{dh}{dt} = -\frac{1}{\tau_h(V)} (h - h_\infty(V)).$$

Here $P_{ca}$ denotes the maximum permeability and *m, h*, denote the activation and inactivation variables, with $\tau_m$ and $\tau_h$ being the corresponding time constants. The kinetic functions are taken from Destexhe et al. (Destexhe et al., 1998). G (*V*, $Ca_o$, $Ca_i$) is a nonlinear function of the potential and ion concentration;

$$G(V,[Ca]_o,[Ca]_i) = \frac{Z^2 F^2 V}{RT} \frac{[Ca]_i - [Ca]_o \exp[-ZFV/RT]}{1 - \exp[-ZFV/RT]}. \tag{2.2.4}$$

Here $Z=2$ denotes the valence of calcium ions, $F$ the Faraday constant, $R$ the gas constant, and $T$ the temperature measured in Kelvin. $[Ca]_i$ and $[Ca]_o$ represent the intracellular and extracellular $Ca^{2+}$ concentrations measured in millimolar. The fluctuation of $Ca^{2+}$ concentration inside the cell denoted by $[Ca]_i$, due to $Ca^{2+}$ pumps and buffers (McCormick and Huguenard, 1992), is taken into account by the following differential equation:

$$\frac{d[C]_i}{dt} = -\frac{i_{ca}}{2Fd} + \frac{([C]_\infty - [C]_i)}{\tau}. \tag{2.2.5}$$

Here $i_{ca}$ denotes the current density, $d = 0.1\mu m$ the depth of the shell below the membrane, $[C]_\infty = 240nM$ the equilibrium concentration and $\tau = 5$ msec the time constant of decay of $[Ca^{2+}]$, respectively. The parameters were taken from the model by Destexhe et al., (Destexhe et al., 1998).

The value of permeability in the low-threshold calcium current (see Eq. 2.2.3) indicates the maximum permeability of the section in unit of cm/sec. This value represents the permeability of a patch of 1 cm$^2$ of membrane. Therefore, the total permeability of the section is obtained by integration over the entire surface. On the other hand, the total permeability of a section is the product of permeability of a single channel and total number of channels. Thus holds

$$N^{total} = \frac{1}{P_{ca}^{(1)}} \int P_{ca}(A) \, dA \tag{2.2.6}$$

Here $N^{total}$, $P_{ca}^{(1)}$ and A denote the total number of T-channels, the permeability of a single channel and surface of compartment, respectively. For a given section area, increasing the number of channels is equivalent to increasing the permeability of sections. In the following, we will refer to the permeability of channels in an area of 1 cm$^2$ of the membrane, simply as channel permeability. In section 3.1., we consider the influence of the number of channels and its distribution on the burst response of TC cell. We also consider in section 3.4 the influence of geometrical parameters (e.g. diameter, area) on the response of the cell.

In order to simplify calculations, we did not include the h-current in the model, because this current does not contribute directly to the generation of LTSs.

## 2.3. Multicompartment model

We developed a multi-compartment model of the TC cell in order to consider the effects of dendritic currents upon response of the cell with realistic morphology. The simulations were performed on a TC cell reconstructed using the NeuroLucida digital system. In order to obtain exact morphological features of the TC neuron from the VPL nucleus of thalamus, a retrograde tracer (fluorogold) was iontophoretically applied to the somatosensory area (SI) in adult anesthetized cat in sterile conditions. After a survival period of two weeks the cat was deeply anesthetized with thiopental and intracardialy perfused. Stained neurons were revealed using standard ABC kit. All experimental procedures used in this study were performed in accordance with the Canadian guidelines for animal care and were approved by the committee for animal care of Laval University. The reconstructed cell used for the model included 11 primary branches and 224 segments with a dendritic membrane area of $45000 \times 10^{-6}$ m$^2$. Our model contains 1267 compartments, based on the d-lambda rule of NEURON. The full compartment model required approximately 12 min on a Pentium(R) M with 2.00 GHz processor to run 800 ms of neuronal activity.

# 3. Results

## 3.1. Calcium channel distribution

In the following simulations, we determined the minimal (threshold) of the number of T-channels ($N^{thr}$) and a corresponding value of permeability ($P^{thr}$) in each section, enabling the generation of LTS in the TC cell. The threshold number of T-channels is defined here as the minimal number of T-channels able to produce LTS in the 3-compartment model. This threshold value corresponds to the onset of the first sodium spike generated by LTS in response to a depolarized current of $I_{inj}$=0.1 nA, while the cell is being hyperpolarized to $V_m$= -75 mV. However, if threshold LTS was obtained under different conditions (e.g., different level of $V_m$ and/or different level of $I_{inj}$) it did not affect the results presented in this study.

## 3.2. Results of numerical simulations

In the first simulation, based on the 3-compartment model, the calcium channel density was kept equal in each compartment, which means equal permeability for $Ca^{2+}$ ions in each compartment. According to experimental data from dissociated cells (Destexhe et al., 1998), the permeability in each compartment was set to the value of $P_{Ca}$=1.7 X $10^{-5}$ cm/sec. Using this value in each compartment, the model generates passive response to a depolarizing injected current (0.1 nA), at a hyperpolarized membrane potential (-75 mV). Parallel increase in permeability in the three compartments by the same amount, induced first a sub-threshold response that grew in amplitude as $P_{Ca}$ increased (Fig. 2 B-D). When the value of permeability at each compartment reached $P_{Ca}$=1.56 X $10^{-4}$ (cm/sec), the modeled cell reproduced first LTS mediated action potential (Fig. 2 E). This permeability was defined as the threshold permeability in conditions of $V_m$=-75mV and $I_{inj}$=0.1nA. The threshold permeability corresponds to the threshold number of $Ca^{2+}$ channels ($N^{thr}$) in the cell requested to generate the first rebound burst in the model. The total number of T-channels in the cell is proportional to multiplication of compartment area to permeability of $Ca^{2+}$ channels in each compartment. The membrane potential fluctuates due to changes in channel permeability, and because the threshold value depends on the level of membrane potential, we used a steady holding current to keep the membrane potential at -75 mV when we changed $P_{Ca}$. In the case of uniform T- channel distributions further increase in permeability produced LTSs of larger amplitude leading to multiple spike generation with increasing intra-burst firing frequency (Fig. 2 F-I). Because the experimental data suggest a non-uniform T-channel distribution in the TC neurons (Zhou et al., 1997, Williams and Stuart, 2000), we examined different shapes of non-uniform distribution of T-channels in the model that lead to LTS response.

The total number of T-channels was kept constant and equal to $N^{thr}$, which is the minimum necessary number of T-channels to generate an LTS response in cells with uniform channel distribution. We explored a variety of T-channel distribution that could reproduce first LTS mediated action potential in the model. Some examples of permeability distribution and corresponding channel number in each compartment leading to an action potential are shown in Figure 3. Both uniform and non-uniform distribution of T-channels can produce LTS response with sodium spike in the modeled cell. This suggests that when the total number of T-channels is equal to $N^{thr}$ then the shape of channel distribution does not affect the LTS response of the cell (Fig. 4 A). However, if the total number of channels was decreased the response of the cell depended on the channel distribution pattern. Far from the threshold value of the total number of T-channels, each form of distributions gave a passive response, but when the total number of T-channels was set close to threshold value, the response of the cell depended on the form of channel distribution (Fig. 4 B, C).

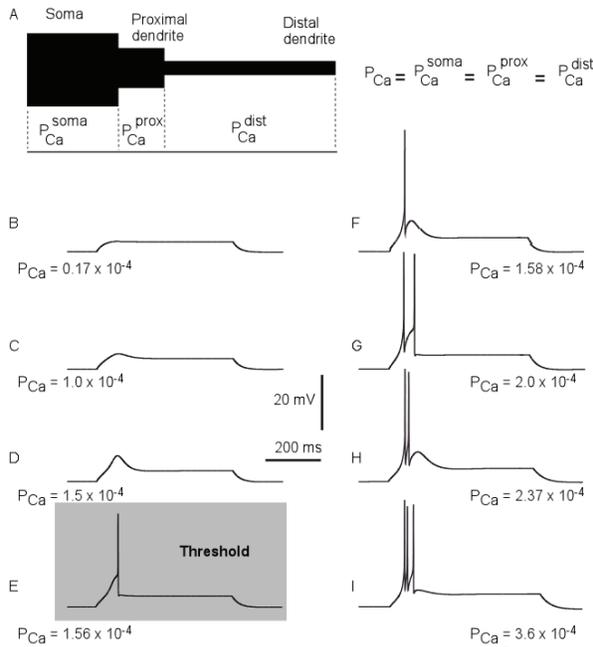

**Figure 2. Effects of permeability for $Ca^{2+}$ increase on generation of LTS responses in a 3-compartment model with uniform channel distribution.**

(A) Schematic representation of 3-compartment model with uniform channel distribution. (B-I) Responses of modeled TC neuron ($V_m$=-75 mV) to depolarizing current pulse (0.1 nA). The permeability for $Ca^{2+}$ in each compartment increases from B to I as indicated. With this morphology, channel distribution and membrane potential the LTS leading to action potential required $P_{ca}$=1.56 x $10^{-4}$ (cm/sec).

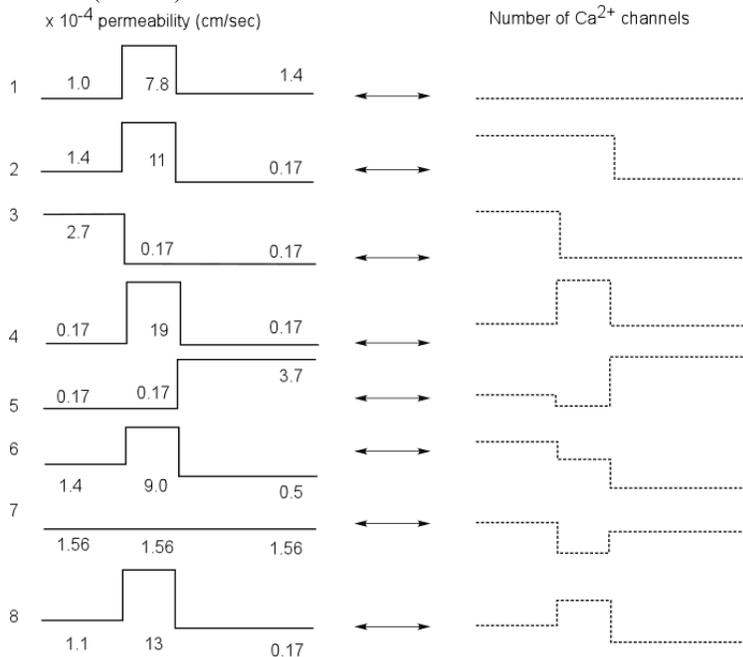

**Figure 3. Different $Ca^{2+}$ channel distribution can generate LTS spiking.**

Each line shows schematically, the level of permeability in each compartment and its corresponding number of channels, which generate threshold sodium spike similar to the response in Fig. 2 E. The channel number in each section is determined by integration over the entire surface of the section. Both uniform and non-uniform distribution of T-channels can produce LTS response in the modeled cell.

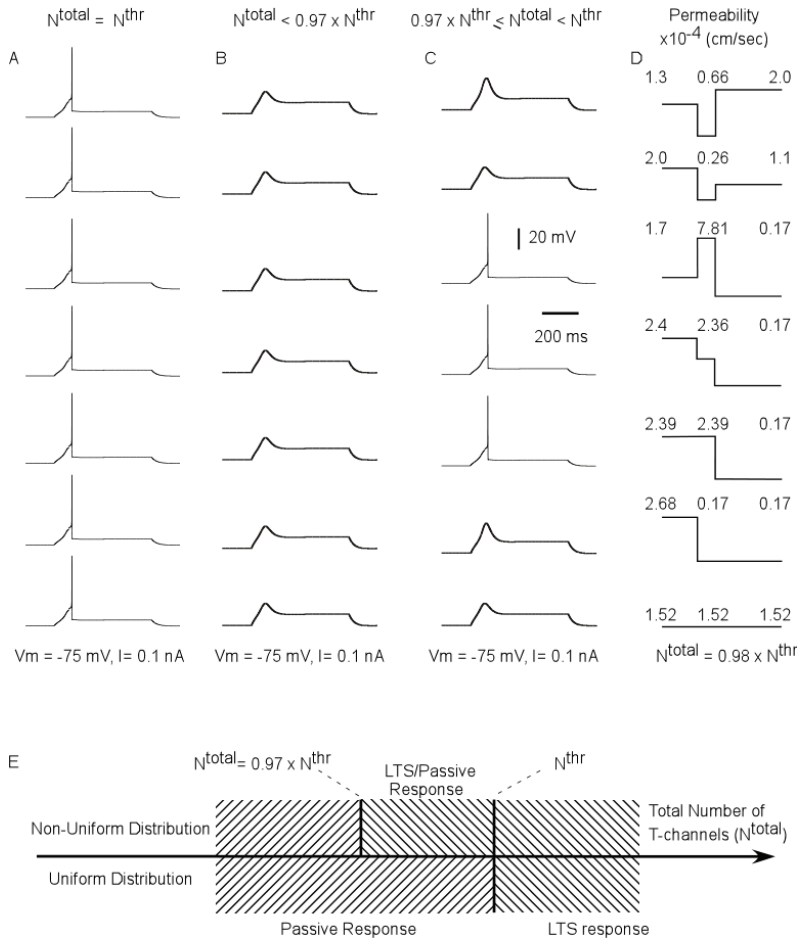

**Figure 4. Total number of Ca²⁺ channels and its distribution determines whether LTS will lead or not lead to action potential.**
(A) When the total number of T-channels was equal to threshold value ($N^{thr}$), the model neuron produced LTS-action potential response with both non-uniform and uniform channel distribution. (B) When the total number of channels was far from threshold value, the model gave LTS response not accompanied with action potentials. In this 3-compartment model, the response has no spikes as long as the total number of T-channel was below 97% of $N^{thr}$. (C) Between these two limits, the response of the neuron was sensitive to the shape of T-channel distribution. To illustrate this feature, we set the total number of channels equal to 98% of $N^{thr}$. (D) Numbers show T-channel permeability in each compartment for different patterns of T-channel distribution, when the total number of T-channels was equal to 98% of $N^{thr}$.

Using small changes in the channel number in the 3-compartment model with non-uniform channel distribution, we found a small window in the total channel number located below the defined threshold of onset of LTS with spikes which was able to induce LTS mediated action potential

$$N_{\min} = 0.97 N^{thr} \leq N^{total} < N_{\max} = N^{thr} \quad . \quad (3.2.1)$$

In such window, the spiking behavior was found to be variable: when the total number of channels was between $N^{thr}$ and 0.97x $N^{thr}$, a uniform permeability distribution did not generated LTS covered with spikes, while a non-uniform distribution could generated full LTS response (Fig. 4). We also tested extreme cases in with Ca²⁺ permeability was 0 in 2 out of 3 compartments, and the $N^{total}$ was still 0.97% of the threshold. These results led to assumption that the modeled cell regenerate rebound burst response for an optimized number of T-channels while those were distributed on a specific non-uniform pattern. We hypothesize that the physiological reason for the existence of such window may be on one side related to reduced metabolic energy consumption (see discussion), and on the other side to ability of the TC cell to generate a variable response pattern.

## 3.3. Effect of cell geometry on firing pattern

Firing patterns do not only depend on the channel density and type of current, but also on the geometry of the cell (length, diameter, area, etc.) (Mainen and Sejnowski, 1996) as well as the cell topology (symmetry, mean path length, etc.) (van Ooyen et al., 2002). Here, we examined the influence of the cell geometry on the firing pattern in the 3-compartment model for TC neuron. For this purpose, we changed the diameter of each compartment, while holding constant the diameter of other section, the channel densities, and the injected current. In addition, we set the total number of channel equal to 98% of $N^{thr}$ where the spiking induced by LTS response of the cell is sensitive to the distribution form of the T-channels. Like in previous simulations, the model includes the $Na^+/K^+$ current in soma and the T-current in soma, proximal and distal compartments. First, a non-uniform T-channel distribution has been considered obeying the following relation:

$$P_{Ca}\ (distal\ dendrite) < P_{Ca}\ (soma) < P_{Ca}\ (proximal\ dendrite)\ .\quad (3.4.1)$$

In control situation, hyperpolarizing current pulse induced LTS response accompanied with action potentials and depolarizing current pulse induced tonic firing with six spikes (Fig. 5 A). We kept constant the channel density through different cell geometry, the $N^{total}$ and $N^{thr}$ varied that altered the response of the cell. With channel density constant, a decrease in the size of all compartments led to an increase in the input resistance, higher frequency of tonic firing and stronger LTS response (Fig. 5 B, I and V). Increasing the size of the cell led to decrease in the input resistance of the cell, and as follow lower excitability, expressed as a reduced firing frequency and no tonic or LTS response (Fig. 5 C, I and V).

We analyzed the affect of size of each section on the cell responses. A reduction in the size of individual compartments (soma, proximal or distal denrites) led to an increase in the input resistance and the frequency of tonic response (Fig. 5 B, II-IV). However, the LTS response to hyperpolarizing current injection was different. When the diameter of soma changed to 1/3 of its initial value, the total number of T-channels reduced to 69% of threshold value, which was far from the lower limit, 0.97%, of total T-channels to generate rebound burst response (Fig. 5, B VI). When the diameter of proximal or distal dendrite was decreased, the $N^{total}$ decreased. Despite that fact, this number was higher than threshold (minimal) number of T-channels required to generate LTS response and during rebound, the neuron fired with large number of action potentials (Fig. 5 B VII-VIII). Increasing size of the cell reduced input resistance resulting in passive responses in the reaction to depolarized and hyperpolarized current (Fig. 5, C I and V). Smaller soma or dendritic section reduced firing frequency and even prevented induction of action potentials during tonic or LTS response (Fig.5, C; II-V, VII and VIII). Again, for a T-current distribution with higher density in proximal section, an LTS response for a bigger soma does not follow the expected pattern. Increasing somatic area reduced tonic frequency but generated more sodium spike during rebound burst. In this case, the total T-channels was 150% more than threshold value, which means the same hyperpolarized current would activate more T-channels that lead to larger LTS response with more action potentials (Fig. 5, C VI). These numerical simulations on a 3-compartment model of TC cell, for a higher density of T-channel in proximal dendritic section, show that smaller distal dendrites and bigger soma are favorable to regenerate rebound burst in TC neuron.

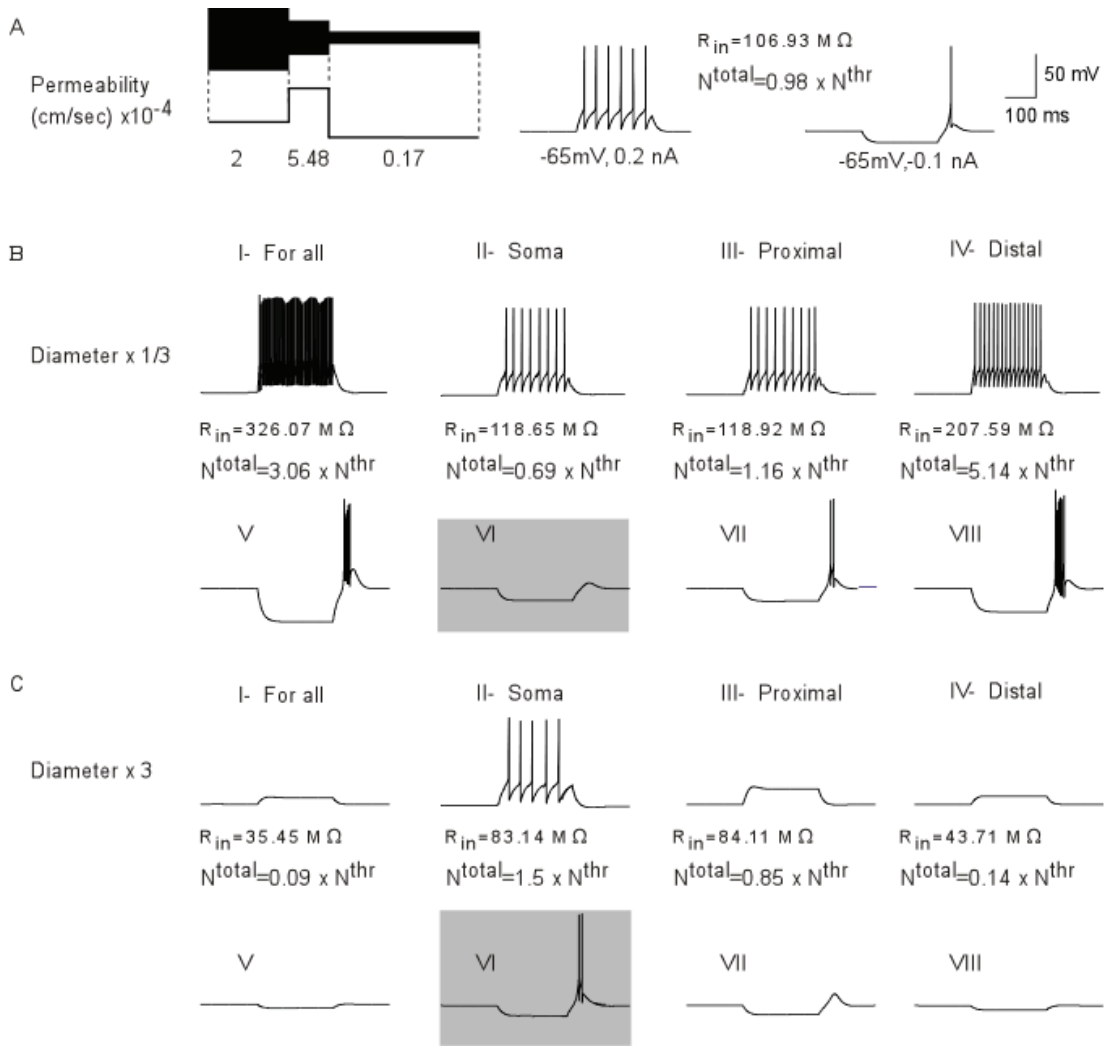

**Figure 5. Influence of geometry on the response of cell with a higher T-channel density in proximal section.**
(A) In control situation ( $N^{total}$ = 98% of $N^{thr}$) the modeled TC cell regenerate 6 spikes in response to 0.2 nA depolarizing current pulse and 1 rebound spike in response to -0.1 nA hyperpolarizing current pulse. (B, C; I - IV) Decreasing/increasing cell size increases/decreases input resistance of the cell that causes higher/lower frequency of tonic response to depolarizing current. (B, C; V, VII, VIII). For a constant density of T-channel, reducing/increasing the size of the cell, or reducing/increasing the dendritic area increased/decreased input resistance as well as $N^{total}$ in the cell that gives more/less sodium spikes induced by LTS. (B, C; IV) Despite the fact that, smaller/bigger soma has higher/lower input resistance, the number of T -channels is also decreased or increased. Therefore, the same hyperpolarized current pulse was sufficient/insufficient to activate enough T-channels for rebound burst response.

Further, we investigated effect of cell size on rebound spike generation when the density of T-current was higher in soma (Fig. 6, A) or distal dendrite (Fig. 6, B). When higher density of T-channels was located on the somatic section, increasing/decreasing somatic diameter increased/reduced the input resistance of the cell and hence, a higher/lower frequency in the responses to depolarizing current pulse was obtained (not shown). When the total numbers of T-channels exceeded the threshold value in bigger cell, which has lower input resistance, the injected hyperpolarizing current activated sufficient number of T-channel to generate rebound burst (Fig. 6, A II and VI). Increasing size of both proximal and distal dendritic section did not shift the ratio of the total and threshold T-channels to a specific range for generation rebound burst response (Fig. 6, A VII and VIII). Decreasing diameter of dendritic sections provided more than threshold value for T-channel that led to more sodium spikes during LTS response (Fig. 6, A III and IV). When higher density of T-channels was located in distal dendrites (Fig. 6, B), reducing the size of the cell increased input resistance and $N^{total}$ that cause more sodium spikes during LTS (Fig. 6, B I-IV). With increased cell size, and constant channel density, the rebound burst was absent (Fig. 6, B V-VIII).

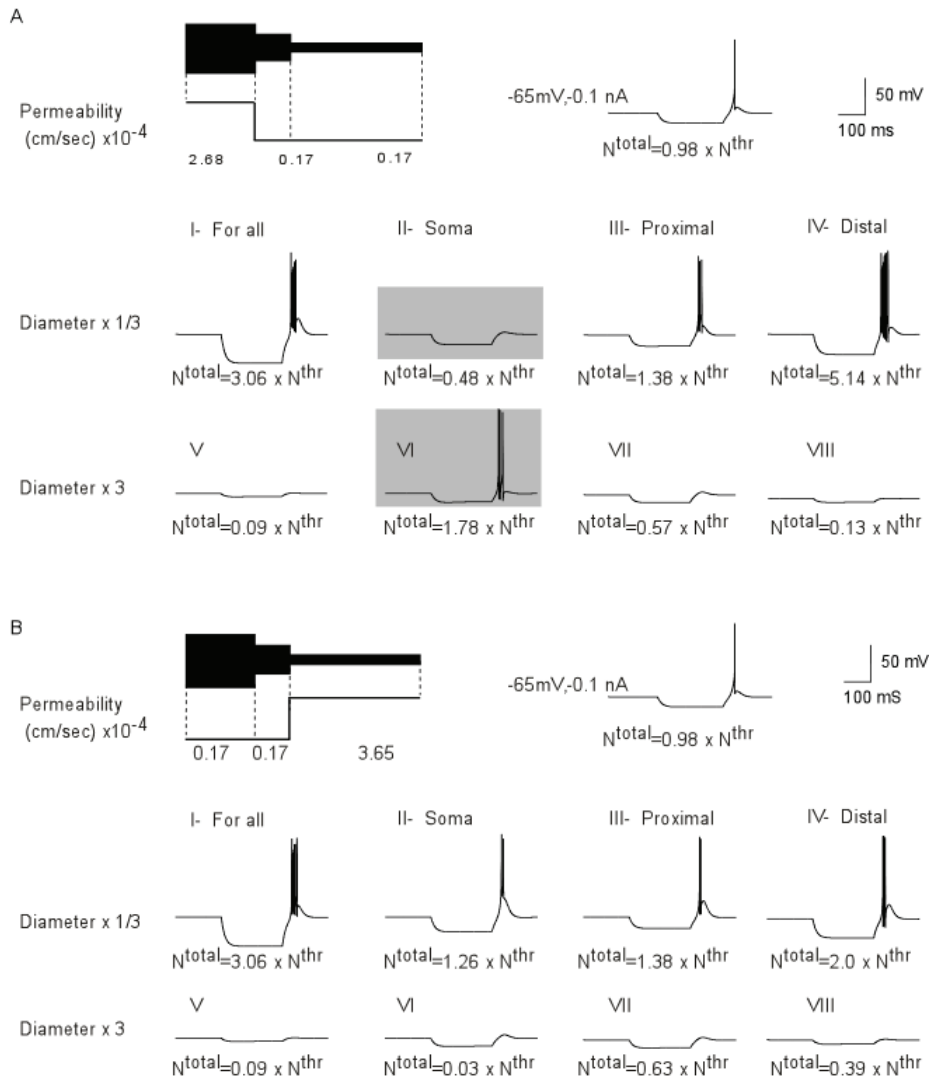

**Figure 6. Influence of geometry on the response of cell with a higher T-channel density in somatic or distal dendritic sections.**

(A) The LTS response with a higher density of T-channels was in soma. (A I, III and IV) Reducing the size of the cell decreased the total number of T-channels, but still this value was higher than threshold, thus model reproduced an LTS response accompanied with burst firing. (A; V, VII, VIII) Increasing the size of the cell increased the total number of T-channels, which was smaller than threshold, thus model reproduced a passive response. (A II) Reducing the size of the soma changes the ratio of total to threshold T-channel to 48%, which was not sufficient to generate LTS. (A VI) increasing the size of soma changed the total number of channel 178%, which leads to rebound burst. (B) The LTS response with a higher density of T-channels was in distal dendrites. (B I- VIII) Reducing cell size set cell more excitable, which reproduces higher frequency for LTS response, while bigger cell has smaller input resistance and the ration of $N^{total}$ to $N^{thr}$ of T channels was outside the range of LTS generation.

## 3.4. Multi-compartmental model.

In a multi-compartment model of the TC cell, reconstructed from fluorogold retrograde staining (Fig. 8, (Ferecskó et al., 2007)), we distributed non-uniformly the T-channels throughout the cell. We assumed a Gaussian distribution with specific mean value attributed to a high density of T-channels in the proximal, middle, and distal dendrites. In addition, the width of the Gaussian function, defined by parameter σ was changed to study the effect of local channel density on LTS generation (Fig. 9). Like in the 3-compartmental model we defined the minimum number of T-channels in the uniform distribution, as the number sufficient to generate an LTS leading to a single spike in response to a depolarized current of 0.1 nA, at hyperpolarized level of Vm=-75 mV. The numerical simulation of LTS response in which a non-uniform T-channel distribution was used shows that a smaller number of T-channels was sufficient to generate an LTS response. In all simulations with non-uniform channels distribution (σ between 0.5 and 2), the LTS with a single spike was generated when $N^{total} = 0.9 \times N^{thr}$, and it was $N^{total} = 0.8 \times N^{thr}$, independently on the location of highest density of T-channels (Fig. 9). However, when the local T-channel density on proximal dendrites was high (σ=0.5), the LTS accompanied by action potential was generated with only $N^{total} = 0.64 \times N^{thr}$ (Fig. 9). Thus, our model predicts that high local channel density on proximal dendrites favor LTS generation with a minimal number of channels.

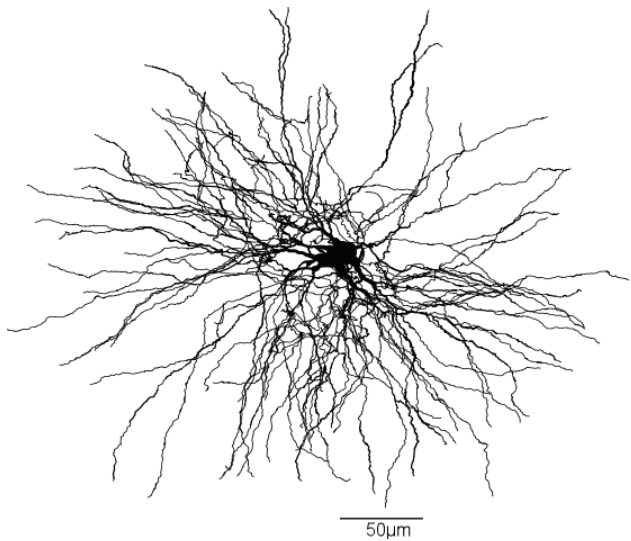

**Figure 7. Morphology of reconstructed TC cell.**
Three-dimensional reconstruction of TC cell from VPL nucleus of adult cat. Retrograde tracer (fluorogold) was injected in somatosensory area SI. The complete dendritic arbor was reconstructed from 6 serial sections 80 μm each.

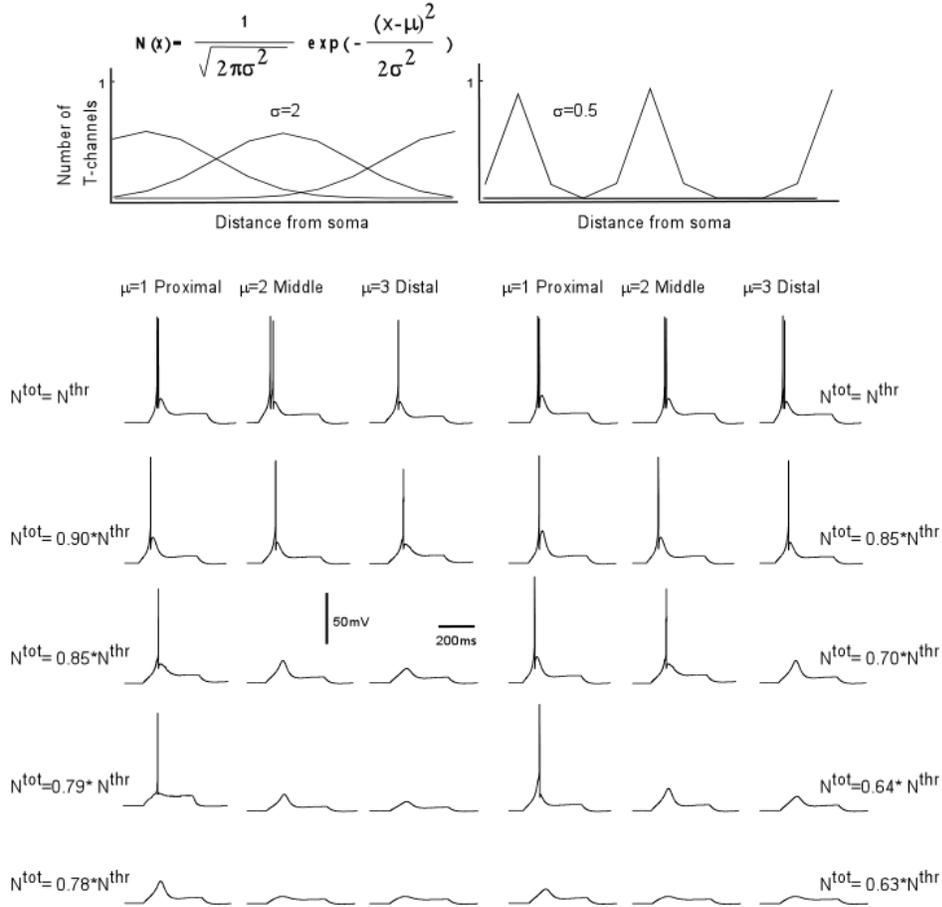

**Figure 8. T-channel distribution required for LTS generation with minimal number of channels in multi-compartmental model.**
Two graphs in the upper panel show the shape of T- channel distribution in the multi-compartment model. In the left column, the maximal channel density slowly decreases from its maximum value. In the right column, the maximal channel density rapidly decreases from its maximum value. Below examples are given for responses of the modeled neuron kept at -75 mV subject to 0.1 nA square current pulses. Note that the lower number of T-channels necessary to generate LTS occurred when the highest channel density was applied in proximal dendrites. In addition, sharp T-channel distribution leads to LTS generation with fewer channels (bottom, right).

## 4. Discussion

The prime objective of this study was to investigate how different patterns of T-channel distributions affect the LTS response of a TC cell in a 3-compartment and multi-compartment model. We also investigated the influence of the cell size on the firing pattern. The major findings of this study are: (i) An increase in the value of permeability (number of channels) uniformly in three compartments changed the response of the TC cell from passive to LTS behavior being accompanied by an action potential. (ii) The onset of firing does not only depend on the total permeability of T-channels, but also on their distribution. (iii) In a 3-compartment model, with a channel number equal or above threshold the TC cell always responded with spikes. If a non-uniform distribution of channels was used, a number of channels smaller than threshold can lead to neuronal firing. (iv) In a multi-compartmental model a minimal number of channels required for the generation of LTS with spikes was much lower than in the 3-compartment model. (v) For a high density of T-channels in the soma and proximal section, the frequency of tonic firing was inversely proportional to the diameter of the section, while the low threshold spiking frequency was directly proportional to the diameter of the section. (vi) The steepness of channel distribution (higher local density of channels) lowers the required minimal number of T-channels necessary for LTS generation.

Previous modeling study on the subject compared results obtained in dissociated cells with cells recorded from 200 μm thick slices (Destexhe et al., 1998). Dissociated cells contained almost no dendrites, while cell recorded from slices contained several truncated dendrites. As we show here, dendritic arbor of thalamocortical cells span out by more than 400 μm (Fig. 7). The study by Destexhe et al. (1998) demonstrated the presence of significant T-current in the dendrites. It also demonstrated that the most robust features of LTS could be reproduced in a reduced 3-compartment model. In our numerical simulations, we took 3-compartment model and demonstrated that in that model the uniform vs. non-uniform channel distribution could affect LTS generation in the range 97-100% (Fig. 4). However, when we used morphologically intact cell to model (Fig. 8), the channel distribution pattern played major role, and if highest density of T-channels was located on proximal dendrites the minimal number of T-channels to generate LTS-spike response was only 64% of threshold. Therefore, comparison of data from the dissociated cells with cells recorded from slices (Destexhe et al., 1998) provided robust differences, because mostly proximal dendrites were present in the cells in slices.

A general principle, that living organisms employ to accomplish many functions, is the use of minimal recourses. The result of our study predicts that similar to sodium channels (Crotty et al., 2006), the non-uniform distribution of T-channels, could be essential for the generation of LTS using minimal resources. Most of proteins, including T-channels are synthesized in the cell body and transported from there to its destination. Protein synthesis and transport require energy. Additionally, the metabolic energy consumption is proportional to the ionic currents generated in the neuron via ATPpase $Na^+/K^+$ or $Ca^{2+}/Na^+$ metabolic exchangers (Laughlin et al., 1998, Goldberg et al., 2003, Lennie, 2003, Dipolo and Beauge, 2006). Following LTS generation, the intracellular $Ca^{2+}$ concentration increases, and reestablishing $Ca^{2+}$ concentration takes additional energy. Thus, optimizing number of channels would optimize metabolic energy consumption. The closest place for T-channels to be inserted in the plasma membrane is the cellular body. However, most of inhibitory synapses are formed on the body of TC neurons (Liu et al., 1995a, Liu et al., 1995b). Thus, the shunting created by inhibitory activities would prevent generation of LTS even with relatively high density of T-channels. The next closest place being energy efficient for T-channels to be integrated in the cell membrane would be proximal portion of the dendritic tree. Indirect experimental measures (Zhou et al., 1997, Williams and Stuart, 2000), congruent with the current study suggested the presence of high density of T-channels on proximal dendrites of the TC cells. Our parallel electron microscopy study using immunogold staining of T-channel subunits in reticular thalamic nucleus demonstrated that in that nucleus, the T-channels are non-uniformly distributed (Kovács et al., 2007). The definite conclusion about the exact distribution of T-channels on the TC cytoplasmic membrane may be reached via use of high-resolution electron microscopy experiments. Optimization of metabolic energy consumption is a decisive factor for the ion channel distribution (Crotty et al., 2006). Our results from the three- and multi-compartment model predict the existence of such window.

Even if T-channels would be equally distributed over the cytoplasmic membrane, the T-channels could be either in the phosphorylated or de-phosphorylated state. Phosphorylation facilitates the generation of T current at resting membrane potential (Leresche et al., 2004). Thus, state dependent changes of the membrane potential could create non-uniform distribution of phosphorylated T-channels in different compartments of TC neurons and would facilitate the generation of LTSs.

We conclude that although different factors could be responsible for a decrease in the availability of T-channels for LTS generation (physical number of channels or ratio of phosphoreleted/de-phosphorelated channels), all the factors inducing a non-uniform distribution of T-channels on LTS generation will favor generation of LTSs. The effect of non-uniform distribution of T-channels on LTS generation in neurons with realistic (much more complex) dendritic tree was found to be stronger.

**Disclosure/Conflict-of-Interest Statement**
The authors declare that the research was conducted in the absence of any commercial or financial relationships that could be construed as a potential conflict of interest.

**Acknowledgements**
H.K has been supported by NSERC Canada. I.T has been supported by CIHR, NSERC Canada and NIH. We are grateful to Alain Destexhe and Maxim Bazhenov for valuable advises that helped to develop our model and K Kovács for help with staining procedure.